\documentclass[journal=jacsat,manuscript=article]{achemso}

\usepackage[version=3]{mhchem} 

\author{Kseniia Storozheva}
 \email{kseniya.storozheva@bisomlab.com}
 \affiliation{ 
    N. N. Semenov Federal Research Center for Chemical Physics Russian Academy of Sciences,  Moscow 119991, Russia}
 \alsoaffiliation{ 
    Design Center for Molecular Machines, Moscow 119991, Russia}
 
\author{Anastasia Markina}
 \email{anastasia.markina@bisomlab.com}
 \affiliation{ 
    N. N. Semenov Federal Research Center for Chemical Physics Russian Academy of Sciences,  Moscow 119991, Russia}
 \alsoaffiliation{ 
    Design Center for Molecular Machines, Moscow 119991, Russia}

\author{Vladik Avetisov}
 \email{avetisov@chph.ras.ru}
 \affiliation{ 
    N. N. Semenov Federal Research Center for Chemical Physics Russian Academy of Sciences,  Moscow 119991, Russia}
 \alsoaffiliation{ 
    Design Center for Molecular Machines, Moscow 119991, Russia}

\title[An \textsf{achemso} demo]
  {Computational Design Rules for Helical Aromatic Foldamers: $\pi-\pi$ Stacking, Solvent Effects, and Conformational Stability}

\abbreviations{IR,NMR,UV}
\keywords{American Chemical Society, \LaTeX}

\begin{document}

\begin{abstract}

Molecular-scale materials with bistable behavior and tunable properties are increasingly relevant for next-generation nanoscale electronic devices. Helical foldamers are promising candidates, but their structural and mechanical properties are highly sensitive to conformational stability and environmental conditions. A systematic methodology based on quantum-chemical calculations is proposed for assessing solvent-dependent mechanical behavior, combining analysis of $\pi-\pi$ stacking interactions, conformational energetics, and environmental effects. Using this methodology we identified simple design principles for the rapid screening of new compounds, allowing evaluation of their conformational stability and effective mechanical rigidity. Applying these principles, we identify a modified helical aromatic foldamer that exhibits improved mechanical and stability characteristics compared to the initial reference compound.

\end{abstract}

\section{Introduction \label{sec:level1}}

In recent decades, the scalability limits of integrated circuits have become a central challenge in electronics, an issue first anticipated by Moore’s law in 1965. While transistor density has historically doubled approximately every two years, continued miniaturization now faces fundamental constraints, including quantum effects, thermal dissipation, and the difficulty of ensuring stable operation at the nanoscale. \cite{limits1,limits2,limits3} To overcome these barriers, nanoscale electronic devices are being actively explored, driving the rapid development of the field of nanoelectronics. This area encompasses a wide range of approaches, including architectures based on quantum dots, carbon nanotubes, and other nanoscale structures. \cite{carbon1,carbon2,carbon3,carbon4}

Within this context, molecular electronics occupies a special place, as individual molecules can act as fundamental building blocks of circuits. \cite{molel1,molel2} Considerable efforts have been directed toward designing molecular sensors, \cite{Shu} switches, \cite{Molmachines} and other devices capable of controlled configurational changes in response to external stimuli. \cite{Aprahamian} A defining feature of such systems is the existence of multiple stable electronic or structural states, enabling switching, information storage, and signal processing at the molecular level. This multistability makes molecular electronics a promising platform for future device technologies.  

Helical oligomers represent a particularly attractive class of nanoscale elements. They combine conformational stability with the ability to undergo controlled self-assembly, and their helical rigidity increases with chain length. \cite{Hindus,sahu2} In addition, they exhibit distinct optoelectronic properties, such as absorption spectra sensitive to molecular size and structure, \cite{Hindus,sahu2} pronounced chirality, \cite{yashima} and tunable electronic characteristics modulated by chain length and chemical composition. \cite{bedi,immanura} Certain systems even support directional charge transport with low attenuation over increasing length, making them strong candidates for molecular diodes and transistors. \cite{yu}

It was shown that short oligomeric nanostructures stabilized by weak non-covalent interactions exhibit bistable dynamics. \cite{shortpyri,NIPMA} This bistability refers to the coexistence of two steady states between which the system can switch under external perturbations such as mechanical loads, electric fields, temperature variations, or thermal noise. Pyridine-pyrrole (PP) and pyridine-furan (PF) nanosprings, in particular, have been shown to exhibit such bistable behavior under mechanical stretching in water and tetrahydrofuran (THF). \cite{SpontVib,shortpyri} Furthermore, when multiple nanosprings are coupled through a common coupling layer, they can display collective bistable dynamics, as demonstrated for pyridine-furan systems coupled by a graphene plate. \cite{collective}

Nevertheless, the current set of studied structures remains relatively limited, which hinders their practical implementation. Expansion of the chemical space of such oligomers could effectively address this limitation through systematic variation of heteroaromatic fragments combined with a detailed analysis of the factors determining their stability and dynamic behavior.

One such factor is the presence of intramolecular $\pi-\pi$ stacking. In the work of Sahu et al., \cite{Hindus} the structure and optoelectronic properties of helical oligomers based on pyridine units and various heterocycles -- furan, pyrrole, and thiophene -- were analyzed in detail. Helix formation was found to be governed by two main mechanisms: geometric asymmetry of the molecular backbone and efficient intercoil $\pi-\pi$ interactions. The study also distinguishes between two conformations of the monomeric unit based on the in-plane orientation of the heterocyclic rings: a cis conformation in which the heteroatoms are located on the same side of the inter-ring bond, and a trans conformation, where they occupy opposite sides. In the cis conformation, the resulting PF-oligomer adopts a more compact geometry, which is initially less stable; however, this trend may reverse as the number of helical turns increases. Intercoil stacking between aromatic fragments enhances molecular rigidity, stabilizes the helical geometry, and promotes electron delocalization along the chain. \cite{huc}

The dynamic behavior of such systems also depends critically on the energetic barriers separating different conformations, such as cis and trans conformations. \cite{ahn} Low energy barriers may lead to frequent spontaneous interconversions at room temperature, causing instability of the device. In contrast, high barriers ensure the stability of the spatial structure, which is necessary for reliable long-term operation. \cite{barriers1,barriers2}

In this study, a model of a molecular spring is examined, whose stability is largely determined by $\pi -\pi$ stacking interactions between aromatic fragments. Particular attention is given to achieving a compact helical folding, which is facilitated by the cis conformation of the monomeric units, ensuring the desired chain curvature. The choice of the cis conformation is further supported by the presence of electrostatic repulsion between heteroatoms, which, in the context of bistable behavior, may facilitate transitions between states. Accordingly, the formulation of the research objectives is based on an analysis of the initial conformation and the factors governing its stability, thereby allowing the identification of key mechanisms underlying the formation of nanosprings in the proposed model.

The present work is aimed at establishing a design riles for the evaluation and selection of heteroaromatic helical foldamers that can serve as the basis for functional elements of nanoelectronics. The study covers the analysis of fundamental factors that determine the structural and mechanical stability of these systems, including the role of $\pi-\pi$ stacking interactions, conformational preferences of monomeric units, and the influence of the external environment, excitation, and protonation. To describe the energy landscapes, density functional theory calculations were carried out on simplified model systems, enabling quantitative characterization of the key interactions that govern the formation of helical conformations.

The obtained results expand the understanding of molecular mechanisms that control the stability and rigidity of bistable helical oligomers and provide a foundation for the rational design of such systems. Based on the performed analysis, a modified chemical structure is proposed with improved characteristics for nanospring configurations.  This study provides a theoretical basis for the development of approaches to optimizing such nanoscale systems, opening avenues for their potential applications in nanoelectronics and other areas of molecular technology.

This article is organized into the following main sections. Section \ref{sec:methods} contains the methods and materials used in this work. Section \ref{sec:results} consists of several subsections. Subsection \ref{susec:solvent} examines $\pi-\pi$ stacking interactions of pyridine-furan in different solvents. Subsection \ref{subsec:rotate} addresses the conformational stability of the monomeric unit, with rotational energy profiles constructed as a function of the dihedral angle for three states: neutral ground, protonated, and electronically excited (first singlet excited state). Finally, subsection \ref{subsec:edot} applies the developed methodology to the evaluation of alternative materials for nanosprings, focusing on the replacement of the furan fragment with an ethylenedioxythiophene (EDOT) unit.

\section{Methods and Materials \label{sec:methods}} 

\subsection{Materials}
\begin{figure}
    \includegraphics[width=0.9\linewidth]{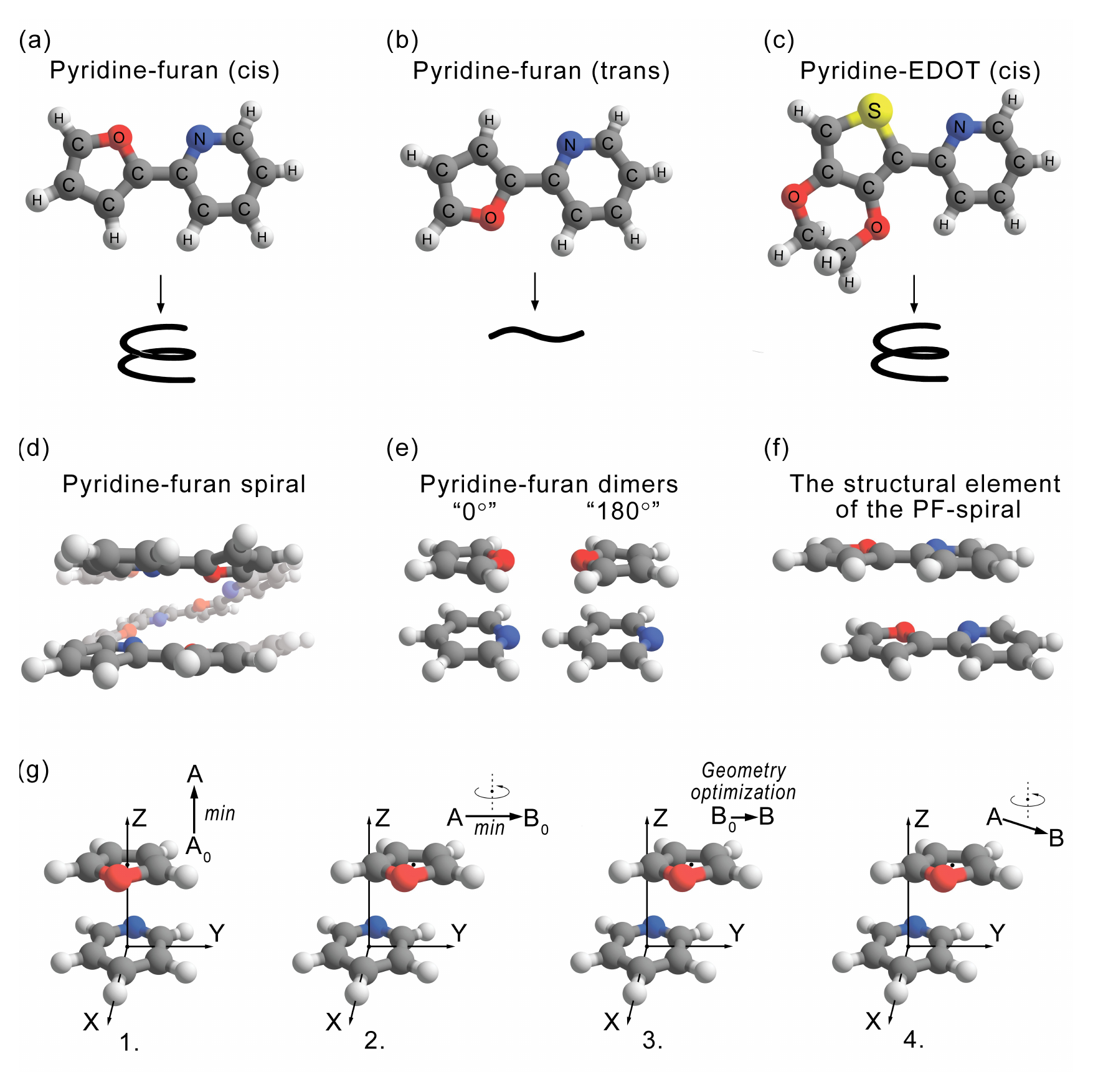}
    \caption{(a) Monomeric unit of pyridine-furan, cis conformation, (b) monomeric unit of pyridine-furan, trans conformation, (c) monomeric unit of pyridine-EDOT, cis conformation, (d) optimized geometry of the pyridine-furan oligomer in THF, front view, (e) pyridine-furan heterodimers in two orientations: $0^\circ$ and $180^\circ$, (f) the structural element of the pyridine-furan spiral, (g) steps of the PES scanning algorithm.}
    \label{fig.ppf5_opt}
\end{figure}

To analyze the potential applicability of helical heteroatomic oligomers as nanoscale functional materials, a pyridine-furan oligomer was chosen, $(-C5H3N-C4H2O-)_{n}$. The monomeric units of this oligomer in the two relevant conformations are shown in Figure~\ref{fig.ppf5_opt} (a) and (b), respectively. This choice was motivated by its chemical simplicity, synthetic accessibility, and ability to form helical structures. Pyridine-ethylenedioxythiophene, $(-C5H3N-C2H4O2C4S-)_{n}$, was chosen as an alternative structure due to its chemical structure and stability. The monomeric unit of this oligomer is shown in Figure~\ref{fig.ppf5_opt} (c).

To select an optimal model for studying $\pi-\pi$ stacking interactions, three structural systems were compared: the spiral oligomer of pyridine-furan (Figure~\ref{fig.ppf5_opt} (d)), a heterodimer of pyridine and furan (Figure~\ref{fig.ppf5_opt} (e)), and the structural element of the oligomer consisting of two monomer units (Figure~\ref{fig.ppf5_opt} (f)). The measurement results of comparing these three structures are provided in the Supporting Information. These calculations demonstrated that the pyridine-furan dimer qualitatively captures the key interactions that govern the properties of the nanospring. This makes it a suitable model system for probing local $\pi-\pi$ stacking interactions.

\subsection{Simulation Details}

To prepare the input data, specifically the initial atomic coordinates, the molecular visualizers Chemcraft \cite{chemcraft} and GaussView~6 \cite{gw} were employed. In these programs, the initial molecular configuration was manually constructed for the first scan point. Subsequent input files for potential energy surface (PES) scanning were then generated automatically using a custom Python script.

\subsection{Energy Surface Scanning Algorithm}

For scanning the PES of the pyridine-furan dimer shown in Figure~\ref{fig.ppf5_opt} (e), we adapted the algorithm proposed by Roland G. Huber et al.~\cite{landscapes} This method enables efficient construction of PES profiles by formalizing the procedure, optimizing the sampling of the energy landscape, and reducing computational cost. Importantly, the sequence of steps reproduces the relative motions of PF-spring monomers observed in molecular dynamics simulations.  

The procedure is illustrated using the pyridine-furan system (Figure~\ref{fig.ppf5_opt}(g)). The coordinate origin is set at the centroid of the pyridine ring, which remains fixed during all calculations. The $Z$ axis is oriented normal to the pyridine plane, and the $X$ axis passes through one of its C-H bonds. The furan fragment is initially placed at the origin of the $X$ and $Y$ axes, with its molecular plane parallel to that of pyridine at a chosen interplanar distance; here it is $Z = 2.9$~\r{A}, see state $A_0$ in Figure~\ref{fig.ppf5_opt}(g).  

Scanning begins by varying the interplanar separation along the $Z$ axis in increments of $0.1$~\r{A}, starting from $Z = 2.9$~\r{A}. The value of $Z$ corresponding to the minimum interaction energy corresponds to the coaxial minimum (state A). Later on in the paper, transition from $A_0$ to $A$ is referred to as a coaxial displacement. From point A, the furan ring is displaced along the $Y$ axis while maintaining a fixed $Z$ coordinate. At each step, it is rotated around the normal to its molecular plane in $30^{\circ}$ increments, covering the full $360^{\circ}$. This identifies the lowest-energy parallel-displaced minimum, termed parallel-displaced stacking; on the Figure~\ref{fig.ppf5_opt}(g) it is referred to as state $B_0$. Geometry optimization is then carried out from this configuration with the pyridine atoms fixed, yielding the ground-state structure (state B). Finally, a trajectory $Y'$ is defined between states A and B, along which the PES is evaluated; this step is referred to as tilted displacement.

\section{Results and Discussion} \label{sec:results}

\subsection{Solvent Effects on the Energy Landscapes of Pyridine-Furan Dimers} \label{susec:solvent}

To gain deeper insight into the structural organization of the molecular nanospring, full geometry optimization was performed for a helical oligomer composed of five monomer units. The optimization was carried out in THF with dielectric constant $\varepsilon=7,43$, selected as a representative weakly polar solvent. The optimized structure is shown in Figure~\ref{fig.ppf5_opt} (d).

The analysis showed that the average interplanar distance between aromatic fragments is $3.45$~\r{A}, which closely matches the value obtained for the pyridine-furan dimer ($3.44$~\r{A}). This also supports the validity of the simplified dimer model in describing key local interactions.
\begin{figure}
    \includegraphics[width=0.6\linewidth]{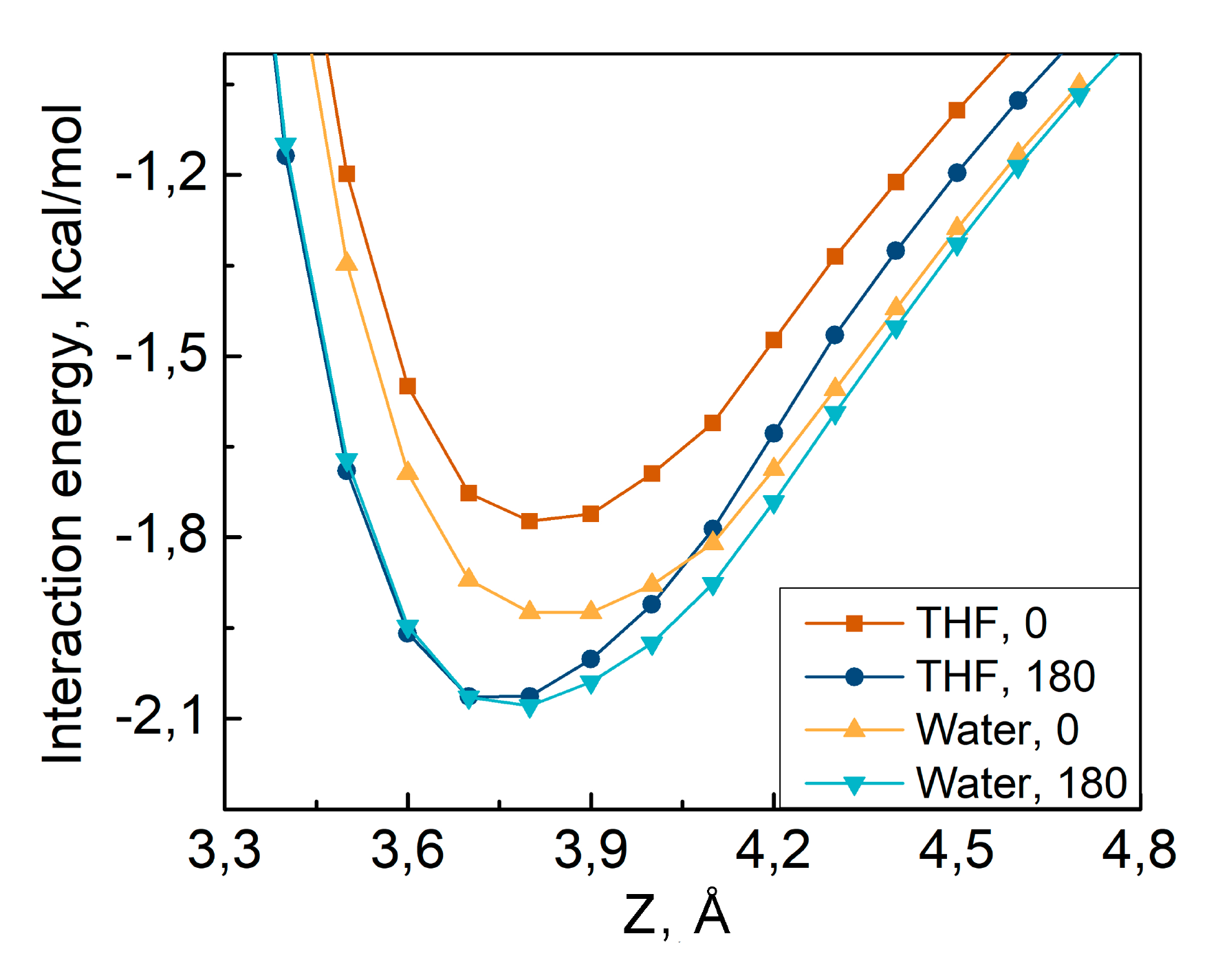}
    \caption{Potential energy profiles of pyridine-furan dimers in $0^\circ$ (orange, yellow) and $180^\circ$ (blue, cyan) orientations in THF and water.}
    \label{fig.tot_graph}
\end{figure}

To explore $\pi-\pi$ stacking energies in more detail, interaction curves were modeled for the coaxial displacement of one monomer relative to the second in two configurations. The first configuration ($0^\circ$ orientation) corresponds to the mutual ring arrangement observed in the optimized nanospring geometry. The second ($180^\circ$ orientation) involves a $180^\circ$ rotation of the furan ring and, as calculations reveal, is energetically more favorable. Both configurations are shown in Figure~\ref{fig.ppf5_opt} (e). The resulting energy profiles for the two solvents, water $(\varepsilon=78,36)$ and THF, are shown in Figure~\ref{fig.tot_graph}.

Figure~\ref{fig.tot_graph} compares the potential energy profiles of pyridine-furan dimers in two relative orientations ($0^\circ$ and $180^\circ$) and in two solvents, water $(\varepsilon=78,36)$ and THF $(\varepsilon=7,43)$. The orange and yellow curves correspond to the $0^\circ$ orientation in THF and water, respectively, while the blue and cyan curves represent the $180^\circ$ orientation in THF and water.

A clear trend is observed when comparing these profiles. The $180^\circ$ orientation (blue and cyan curves) consistently lies below the corresponding $0^\circ$ orientation (orange and yellow curves), indicating a deeper potential well and greater stability. In contrast, the $0^\circ$ orientation shows shallower minima shifted along the horizontal axis, which reflects steric repulsion between the closely positioned nitrogen and oxygen atoms. This repulsion destabilizes the complex and raises the interaction energy, as seen from the higher position of the orange and yellow curves.  

The solvent also plays a decisive role. The difference between the two orientations is more pronounced in THF (orange-blue pair) than in water (yellow-cyan pair). This indicates that orientational sensitivity is greater in a low-dielectric environment, where electrostatic repulsion is less effectively screened. In water, with its high dielectric constant ($\varepsilon$), the repulsion is strongly reduced, leading to smaller differences between the curves.  

These results highlight that the relative orientation of the monomer units critically determines nanospring stability: the $0^\circ$ configuration, though less favorable energetically, is essential for maintaining the compact helical geometry and bistable behavior. At the same time, the dielectric constant of the medium serves as a reliable predictor of how solvent polarity modulates the balance between steric repulsion and $\pi-\pi$ stabilization.

Following the analysis of how monomer orientation affects the stability of the complex, the next step involved a detailed examination of the PES of the pyridine-furan dimer. Given the complex nature of the nanospring’s bistable motion, a PES scanning algorithm was employed to capture the key features of the relative displacement between monomer units. The results obtained at all stages of this procedure are provided in the Supporting Information. This analysis enabled a characterization of the dimer’s energy landscape, including the estimation of the $\pi-\pi$ stacking distance and the location of the corresponding minimum. In addition, the role of the solvent in shaping the stacking interactions was examined by comparing the results of the first two stages of the algorithm - coaxial and parallel-displaced stacking. This approach enables the monomers to follow identical trajectories within the region of interest, thereby isolating the solvent’s effect on the shape of the energy landscape.

To characterize the potential energy landscape in different solvents, a scan of the parallel displacement of furan relative to pyridine was performed at a fixed Z-coordinate, $Z = 3.8$ ~\r{A}. The solvents selected span a wide range of dielectric constants: hexane $(\varepsilon = 1.88)$, benzene $(\varepsilon = 2.27)$, THF $(\varepsilon = 7.43)$, pyridine $(\varepsilon = 12.98)$, methanol $(\varepsilon = 32.61)$, and water $(\varepsilon = 78.36)$.
\begin{figure}
    \includegraphics[width=1\linewidth]{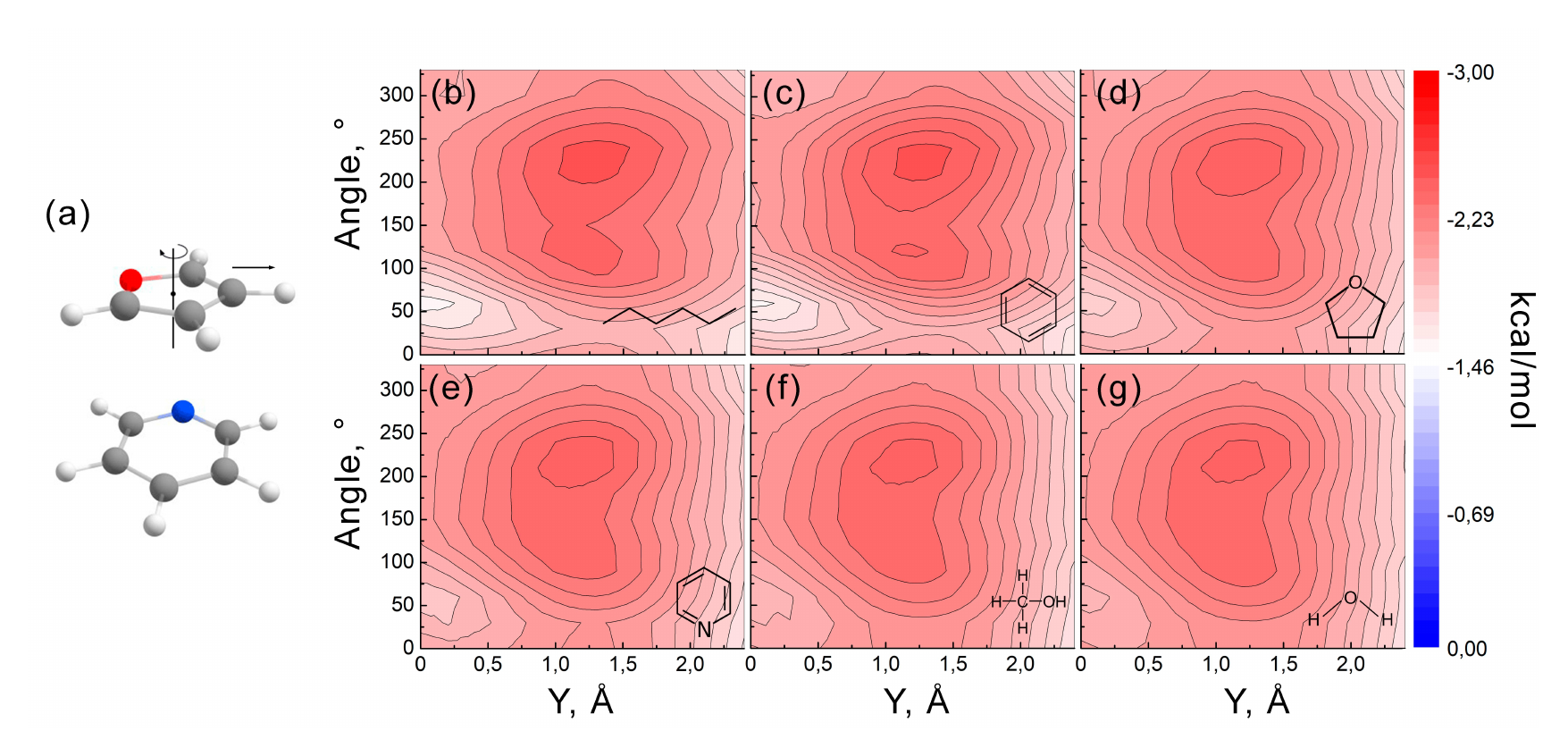}
    \caption{(a) Initial configuration at $0^\circ$ angle (step 2 of the scanning algorithm: parallel displacement and rotation). PES for the parallel-displaced geometry of furan relative to pyridine in various solvents at the constant coordinate $Z = 3.8 $~\r{A}:  (b) hexane, (c) benzene, (d) THF, (e) pyridine, (f) methanol, (g) water.}
    \label{fig.pyri-fur_pes_tot}
\end{figure}

Figure~\ref{fig.pyri-fur_pes_tot} shows the potential energy surfaces for the parallel-displaced stacking geometry in the listed solvents. Analysis of these results demonstrates that increasing the dielectric constant leads to a reduction in contrast between energy minima and maxima.
\begin{figure}
    \includegraphics[width=0.7\linewidth]{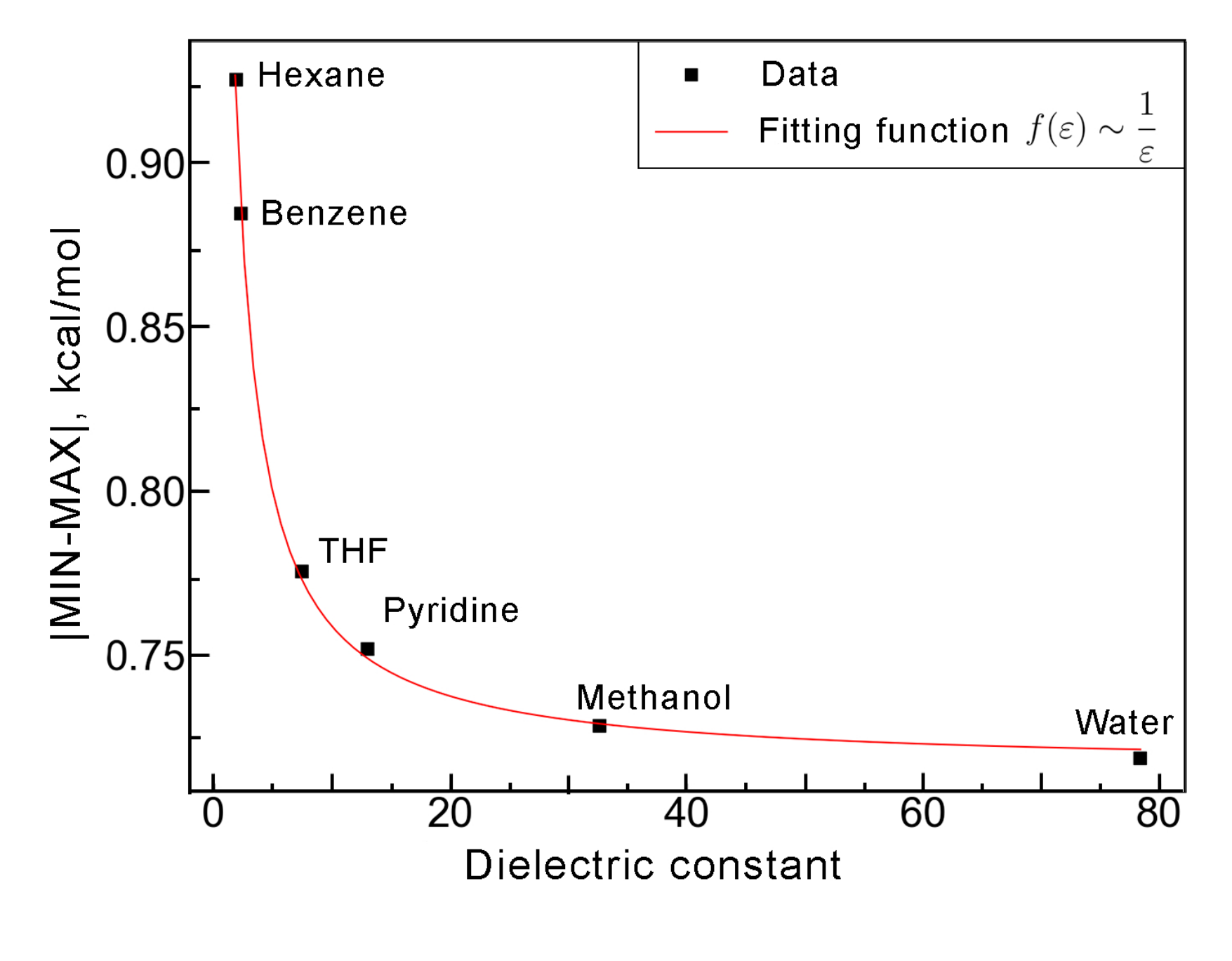}
    \caption{Dependence of the energy interaction difference between the maximum and minimum on the solvent dielectric constant, with fitted approximation.}
    \label{fig.diel_pronic}
\end{figure}

To quantitatively assess the influence of dielectric constant on the energy profile, energy values at local minima and maxima were extracted from the landscapes in Figure~\ref{fig.pyri-fur_pes_tot}. Based on these values, a plot was constructed showing the absolute energy difference between the maximum and minimum as a function of solvent dielectric constant (Figure~\ref{fig.diel_pronic}). This approach provides a numerical estimate of the sensitivity of intermolecular interactions to the polarity of the surrounding medium.

As seen from the plot, the energy difference decreases monotonically with increasing $\varepsilon$, reflecting a weakening of the interaction between the aromatic fragments. The resulting trend was fitted using the function:
\begin{equation}
    f(\varepsilon) = \frac{B}{A + \varepsilon} + C ,
\end{equation}
which reflects the expected form of Coulombic interaction between point charges in a dielectric medium. At large values of $\varepsilon$, the function asymptotically approaches the constant $C$, which accounts for residual van der Waals interactions and dispersion.

Thus, this analysis shows that the dielectric constant of the environment is a key parameter shaping the potential energy landscape of the pyridine-furan system. These findings are important for understanding the role of $\pi-\pi$ stacking in stabilizing helical aromatic foldamers, as well as for selecting appropriate solvents and parameterizing molecular interaction models relevant to the design of nanospring-based materials.

\subsection{Conformational Stability of the Pyridine-Furan Monomer Unit} \label{subsec:rotate}
\begin{figure}
    \includegraphics[width=0.8\linewidth]{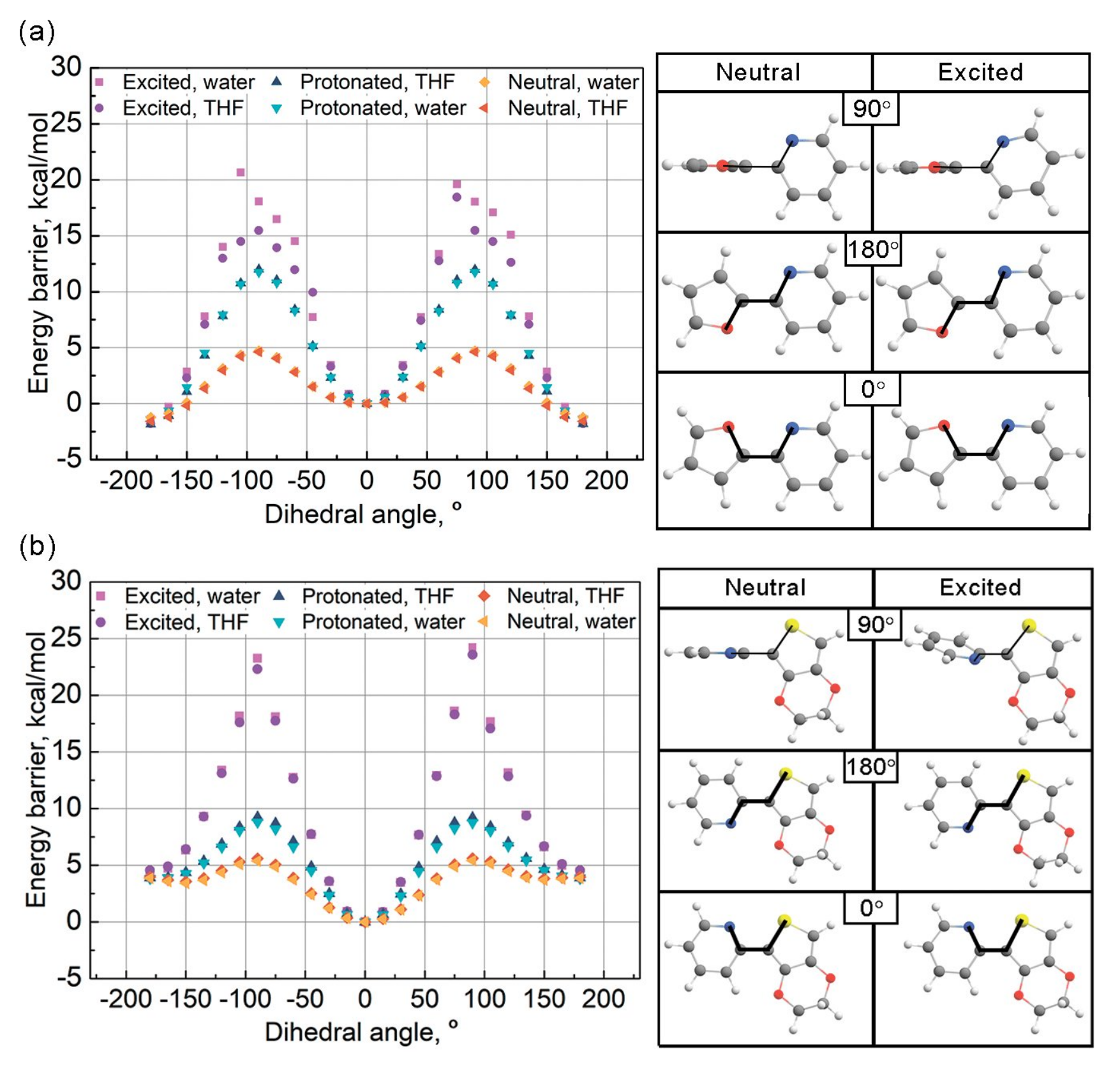}
    \caption{(a) Left: rotational energy profiles of the pyridine-furan monomeric unit around the dihedral angle in neutral, protonated, and excited states in water and THF. Right: molecular geometries at various dihedral angles in the neutral and excited states. The atoms forming the studied dihedral angle are connected by black lines. (b)~Left: rotational energy profiles of the pyridine-EDOT monomeric unit around the dihedral angle in neutral, protonated, and excited states in water and THF. Right: molecular geometries at various dihedral angles in the neutral and excited states. The atoms forming the studied dihedral angle are connected by black lines.}
    \label{fig.pyri-fur_dih-p}
\end{figure}

To assess the influence of solvent on conformational behavior, two media were considered: water and THF. The energy profiles were computed as functions of the dihedral angle defined by the atoms connected by bold lines in the structural scheme shown in Figure~\ref{fig.pyri-fur_dih-p} (a). All energies are normalized to the conformation with a $0^\circ$ dihedral angle.

As shown in Figure~\ref{fig.pyri-fur_dih-p} (a), in the neutral state the rotational energy barrier is approximately 4-5 kcal/mol, exceeding the magnitude of thermal fluctuations at room temperature ($RT \approx 0.593$ kcal/mol). Nevertheless, this does not entirely preclude thermally induced cis-trans transitions at $T = 298$~K. Protonation of the molecule raises the barrier by over 7 kcal/mol, suggesting enhanced rigidity and a reduced likelihood of conformation transitions at room temperature.

The most pronounced effect is observed in the excited state, where the energy barrier exceeds 15 kcal/mol. Additionally, structural distortions appear at dihedral angles of $75^\circ$-$90^\circ$, likely caused by the redistribution of electron density in the excited state, which disrupts the $\pi$-conjugation between pyridine and furan. When the two rings become orthogonal, their $\pi$-systems cease to overlap, allowing heteroatoms to slightly shift into energetically favorable positions, even if this leads to mild out-of-plane distortions.

The analysis of the conformational stability of the pyridine-furan-based monomeric unit showed that the selected orientation underlying the nanospring design is metastable. Although an energy barrier preventing free rotation is present, its magnitude in the neutral state only slightly exceeds thermal fluctuations at room temperature. This suggests that thermally induced transitions to alternative conformations may occur even before the onset of helical folding. It is important to note that once the helix is formed, the orientation of the fragments becomes additionally stabilized by intermolecular interactions, significantly reducing the probability of conformational transitions. However, the initiation of coiling requires the presence of a cis conformation, making it critical for the formation of an ordered structure. Thus, the initial metastability can hinder helix formation, even though the final structure is stable.

\subsection{Screening of a New Material for Molecular Nanosprings} \label{subsec:edot}

To overcome the metastability of the monomer unit, an alternative monomer structure was proposed, one that inherently adopts a stable cis orientation. This system is a pyridine-ethylenedioxythiophene dimer, \cite{edot} Figure~\ref{fig.ppf5_opt} (c). A key structural feature of this compound is that the target orientation corresponds to the global minimum of the potential energy surface, making it energetically favorable from the outset. As in the previous case, dihedral energy scans were carried out for the neutral ground, protonated, and first excited states. The results are shown in Figure~\ref{fig.pyri-fur_dih-p} (b).

In the neutral state, the energy barrier is about 5-6 kcal/mol, providing slightly greater conformational stability than in the pyridine-furan case. Protonation increases the barrier to 7-8 kcal/mol, further stabilizing the desired orientation. The effect is most pronounced in the excited state, where the barrier of 22-25 kcal/mol greatly reduces the likelihood of cis-trans transitions at room temperature. As in the case of pyridine-furan, geometrical distortions are observed at certain dihedral angles.

Thus, switching to pyridine-EDOT provides not only conformational stability in the ground state but also improved resistance to external perturbations. This makes the new system more suitable for use as a tunable molecular spring.
\begin{figure}
    \includegraphics[width=1\linewidth]{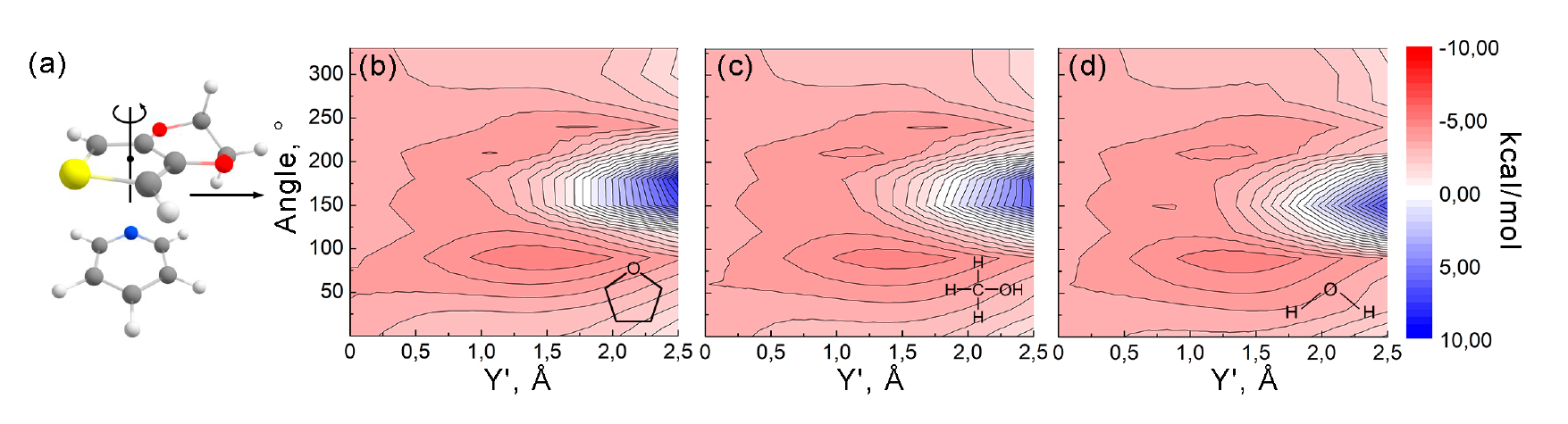}
    \caption{(a) The initial configuration at $0^\circ$ angle  (step 5 of the scanning algorithm: tilted displacement). Next, the potential energy surfaces of the pyridine-EDOT dimer in various solvents: (b) THF, (c) methanol, (d) water.}
    \label{fig.ethyl_pes}
\end{figure}

The analysis of the PES for the new pyridine-EDOT-based structure was carried out using the same methodology as for pyridine-furan. However, in this case, all calculations were carried out for the cis conformation, with the monomers arranged such that the thiophene ring is positioned above the pyridine ring. Energy density maps were used to assess the strength of attraction between fragments and to identify stable regions suitable for designing molecular devices with customized mechanical properties. Intermediate results are available in the Supporting Information; the final PESs are shown in Figure~\ref{fig.ethyl_pes}.

The resulting PES maps exhibit behavior largely similar to the pyridine-furan system. In particular, the potential wells in less polar solvents appear narrower and deeper, consistent with the theoretical prediction that lower dielectric constants enhance structural stability. Another important observation is the presence of regions with positive interaction energies, indicating significant repulsion between the monomers in certain configurations. The appearance of such energetically unfavorable zones can be considered a beneficial factor, as they restrict the accessible conformational space and contribute to the overall rigidity of the system.

\section*{Conclusions}

In this work, we developed a methodology for the systematic assessment of helical aromatic foldamers stabilized by $\pi-\pi$ stacking interactions. The approach integrates three key aspects: (i) intercoil interactions and stacking energetics, (ii) solvent effects on the potential energy surface, and (iii) conformational stability of monomeric units. Together, these elements form a general framework for evaluating and optimizing new candidate structures.  

A central outcome of this study is the identification of the main factors governing the stabilization of helical aromatic foldamers. The dielectric constant of the environment was shown to strongly modulate $\pi-\pi$ stacking interactions, with an analytical approximation derived to capture solvent dependence without additional quantum-chemical calculations. The relative orientation of heteroatoms (cis vs.\ trans) emerged as a decisive structural parameter, directly influencing helical folding and stability through electrostatic interactions. Furthermore, rotational energy scans revealed that while thermal fluctuations may induce conformational transitions in neutral monomers, protonation and electronic excitation substantially increase rigidity, suggesting possible external control mechanisms for device applications.  

Applying this framework, we consider pyridine-ethylenedioxythiophene (EDOT) architecture that retains the favorable features of the pyridine-furan system while providing superior conformational stability. This demonstrates the utility of the methodology as a predictive tool for scanning and designing molecular architectures with tailored bistability and rigidity for future nanoscale electronic applications. 

\section*{Supporting Information Description}

Additional computational details, solvent-dependent potential energy surfaces, and interaction energy profiles for the pyridine-furan, pyridine-EDOT, and nanospring monomer units are provided in the Supporting Information (PDF).

\begin{acknowledgement}

The authors thank Vladimir Bochenkov for insightful discussions regarding the design and architectures of bistable oligomers. 

\end{acknowledgement}

\providecommand{\latin}[1]{#1}
\makeatletter
\providecommand{\doi}
  {\begingroup\let\do\@makeother\dospecials
  \catcode`\{=1 \catcode`\}=2 \doi@aux}
\providecommand{\doi@aux}[1]{\endgroup\texttt{#1}}
\makeatother
\providecommand*\mcitethebibliography{\thebibliography}
\csname @ifundefined\endcsname{endmcitethebibliography}  {\let\endmcitethebibliography\endthebibliography}{}

\end{document}